\documentclass[aps,10pt,pra,superscriptaddress,showpacs,floatfix,twocolumn,longbibliography,nofootinbib,pdftex]{revtex4-1}

\usepackage{graphicx}
\usepackage{epstopdf}
\usepackage{amsmath}
\usepackage{comment}
\usepackage{amssymb}
\usepackage{hyperref}
\usepackage{bm}
\usepackage{subfigure}
\usepackage{float} % loaded by algorithm, doesn't work with Revtex, so use [H] flag to prevent float
\usepackage{algorithm}
\usepackage{algpseudocode}
\usepackage{colortbl}
\usepackage{rotating}
\usepackage{longtable}
\usepackage{enumitem}
\setlist{nosep}

\hypersetup{
    colorlinks=true,       % false: boxed links; true: colored links
    linkcolor=cyan,          % color of internal links
    citecolor=magenta,        % color of links to bibliography
    filecolor=magenta,      % color of file links
    urlcolor=cyan,           % color of external links
    runcolor=cyan
}
\usepackage[pdftex]{color}

\usepackage{threeparttable}
\usepackage{subfigure}
\usepackage{upgreek }
\usepackage{marginnote}
\usepackage{cancel}

\newcommand{\beq}{\begin{equation}}
\newcommand{\eeq}{\end{equation}}
\newcommand{\beqnn}{\begin{equation*}}
\newcommand{\eeqnn}{\end{equation*}}
\newcommand{\bea}{\begin{eqnarray}}
\newcommand{\eea}{\end{eqnarray}}
\newcommand{\beann}{\begin{eqnarray*}}
\newcommand{\eeann}{\end{eqnarray*}}
\newcommand{\bes} {\begin{subequations}}
\newcommand{\ees} {\end{subequations}}

\newcommand{\ignore}[1]{}

\begin{document}
%\title{Scaling Benchmark of Quadratic Unconstrained Binary Optimizers} 
\title{3-Regular 3-XORSAT Planted Solutions Benchmark of Classical and Quantum Heuristic Optimizers} 

\author{Matthew Kowalsky}
\affiliation{Department of Physics and Astronomy, University of Southern California, Los Angeles, California 90089, USA}
\affiliation{Center for Quantum Information Science \& Technology, University of Southern California, Los Angeles, California 90089, USA}
\affiliation{USRA Research Institute for Advanced Computer Science,615 National, Mountain View, California 94043, USA}
\affiliation{Quantum AI Laboratory (QuAIL) NASA Ames Research Center,Mail Stop 269-1, Moffett Field, California 94035, USA}

\author{Tameem Albash}
\affiliation{Department of Electrical and Computer Engineering,  University of New Mexico, Albuquerque, New Mexico 87131, USA}
\affiliation{Department of Physics and Astronomy and Center for Quantum Information and Control, CQuIC, University of New Mexico, Albuquerque, New Mexico 87131, USA}

\author{Itay Hen}
\affiliation{Department of Physics and Astronomy, University of Southern California, Los Angeles, California 90089, USA}
\affiliation{Center for Quantum Information Science \& Technology, University of Southern California, Los Angeles, California 90089, USA}
\affiliation{Information Sciences Institute, University of Southern California, Marina del Rey, California 90292, USA}

\author{Daniel A. Lidar}
\affiliation{Department of Physics and Astronomy, University of Southern California, Los Angeles, California 90089, USA}
\affiliation{Center for Quantum Information Science \& Technology, University of Southern California, Los Angeles, California 90089, USA}
\affiliation{Department of Electrical and Computer Engineering, University of Southern California, Los Angeles, California 90089, USA}
\affiliation{Department of Chemistry, University of Southern California, Los Angeles, California 90089, USA}

\begin{abstract}
With current semiconductor technology reaching its physical limits, special-purpose hardware has emerged as an option to tackle specific computing-intensive challenges. Optimization in the form of solving Quadratic Unconstrained Binary Optimization (QUBO) problems, or equivalently Ising spin glasses, has been the focus of several new dedicated hardware platforms.  These platforms come in many different flavors, from highly-efficient hardware implementations on digital-logic of established algorithms to proposals of analog hardware implementing new algorithms. In this work, we use a mapping of a specific class of linear equations whose solutions can be found efficiently, to a hard constraint satisfaction problem (3-regular 3-XORSAT, or an Ising spin glass) with a `golf-course' shaped energy landscape, to benchmark several of these different approaches. We perform a scaling and prefactor analysis of the performance of Fujitsu's Digital Annealer Unit (DAU), the D-Wave Advantage quantum annealer, a Virtual MemComputing Machine, Toshiba's Simulated Bifurcation Machine (SBM), the SATonGPU algorithm from Bernaschi \textit{et al.}, and our implementation of parallel tempering. We identify the SATonGPU and DAU as currently having the smallest scaling exponent for this benchmark, with SATonGPU having a small scaling advantage and in addition having by far the smallest prefactor thanks to its use of massive parallelism. 
Our work provides an objective assessment and a snapshot of the promise and limitations of dedicated optimization hardware relative to a particular class of optimization problems.
\end{abstract}

\maketitle
%%%%%%%%%%%%%%%%%%%%%%%%%%%%%%%

Many problems of theoretical and practical relevance can be cast as combinatorial optimization problems over discrete configuration spaces. Of particular interest are Quadratic Unconstrained Binary Optimization (QUBO) problems, appearing in diverse fields such as machine learning, materials design, software verification, flight-traffic control, constraint satisfaction problems, portfolio management, and logistics, to name a few examples~\cite{Wol1999,Pap2013}. These problems can be cast as finding the optimal $n$-bit configuration that minimizes 
the cost function
\beq
f_Q(x) = \sum_{i\le j}^n Q_{ij} x_i x_j \ ,
\label{eq:cost}
\eeq
with $x_i \in \left\{0,1 \right\}$. Equivalently, this problem can be cast as finding the ground state of an Ising Hamiltonian, with the bits $x_i$ mapped to spin variables $s_i \in \left\{-1, 1\right\}$ via $s_i = 1-2x_i$, after which the cost function becomes $\sum_i h_{i}s_{i} + \sum_{i < j} J_{ij}s_i s_j$.  The non-zero off-diagonal elements of $Q_{ij}$ define the weighted connectivity graph of a QUBO instance, and the diagonal elements $Q_{ii}$ define its effective `local fields'. Likewise, in the Ising formulation the couplings $J_{ij}$ and local fields $h_i$ define an Ising problem instance.
QUBO is an NP-hard problem, so it is expected that all general-purpose QUBO solvers will have an exponential runtime-scaling with problem size $n$, making solving such problems at large $n$ computationally intractable.

The ubiquity of QUBO problems has led to considerable effort being devoted to reducing and mitigating the computational effort needed to tackle them. For example, specialized hardware implementing FPGA-based digital annealers~\cite{Matsubara2017,Tsukamoto2017}, quantum annealers based on superconducting flux qubits~\cite{Kaminsky-Lloyd,Joh2010,Ber2010,Har2010,Dwave,Bun2014,Weber:2017aa,Novikov:2018aa,grover2020fast}, and coherent Ising machines based on lasers and degenerate optical parametric oscillators~\cite{McM2016,Yam2017,ng2021efficient} have been realized. Similarly, proposals for memcomputing devices that operate on terminal-agnostic self-organizing logic gates~\cite{Tra2015,Tra2018,Div2018} and FPGA-based annealers that use stochastic cellular automata to perform spin updates in parallel~\cite{Yam2021} have been made. 

Published work on these novel solvers features an array of benchmarks~\cite{Ron2014,Hen2015,2016arXiv160401746M,Alb2018,Ham2019,Ara2019,Mat2020,Got2019,Ham2019,Per2019,Sek2020,Aik2020,king2020performance,Goto:2021tb,calaza2021garden}.
Here, we perform a scaling analysis of the time-to-solution (TTS) for a class of QUBO problems designed to provide a characterization of the current and expected {future} performance of some of these solvers. Our benchmark problems have the advantage of a known optimal solution by construction, and the resulting mapping to QUBO retains the hardness properties of spin-glasses when solved using heuristic solvers. The QUBO solvers benchmarked here include: the Virtual MemComputing Machine~\cite{Aik2020}, Toshiba's Simulated Bifurcation Machine~\cite{Got2019b}, Fujitsu's Digital Annealer Unit~\cite{Mat2020}, the D-Wave Advantage quantum annealer~\cite{boothby2020nextgeneration}, a SAT algorithm from Bernaschi \textit{et al.} called SATonGPU~\cite{bernaschi2021leading}, and our implementation of parallel tempering~\cite{Swe1986,Gey1991,Huk1996}. 

All the solvers studied here exhibit an exponential scaling with problem size, with varying degrees of constant-factor overheads. 
We find small variations in the scaling exponent among most of the solvers (the exception being D-Wave, whose exponent is significantly larger), and large variations in the prefactor. Before giving the details of our findings, we describe our problem instances, metric, and briefly review the different solvers. We conclude with a discussion of our results and provide additional details on the solvers in the Methods section. The Appendix provides further technical details and complementary plots.

\begin{figure}[thbp]
    \centering
    \includegraphics[width=\columnwidth]{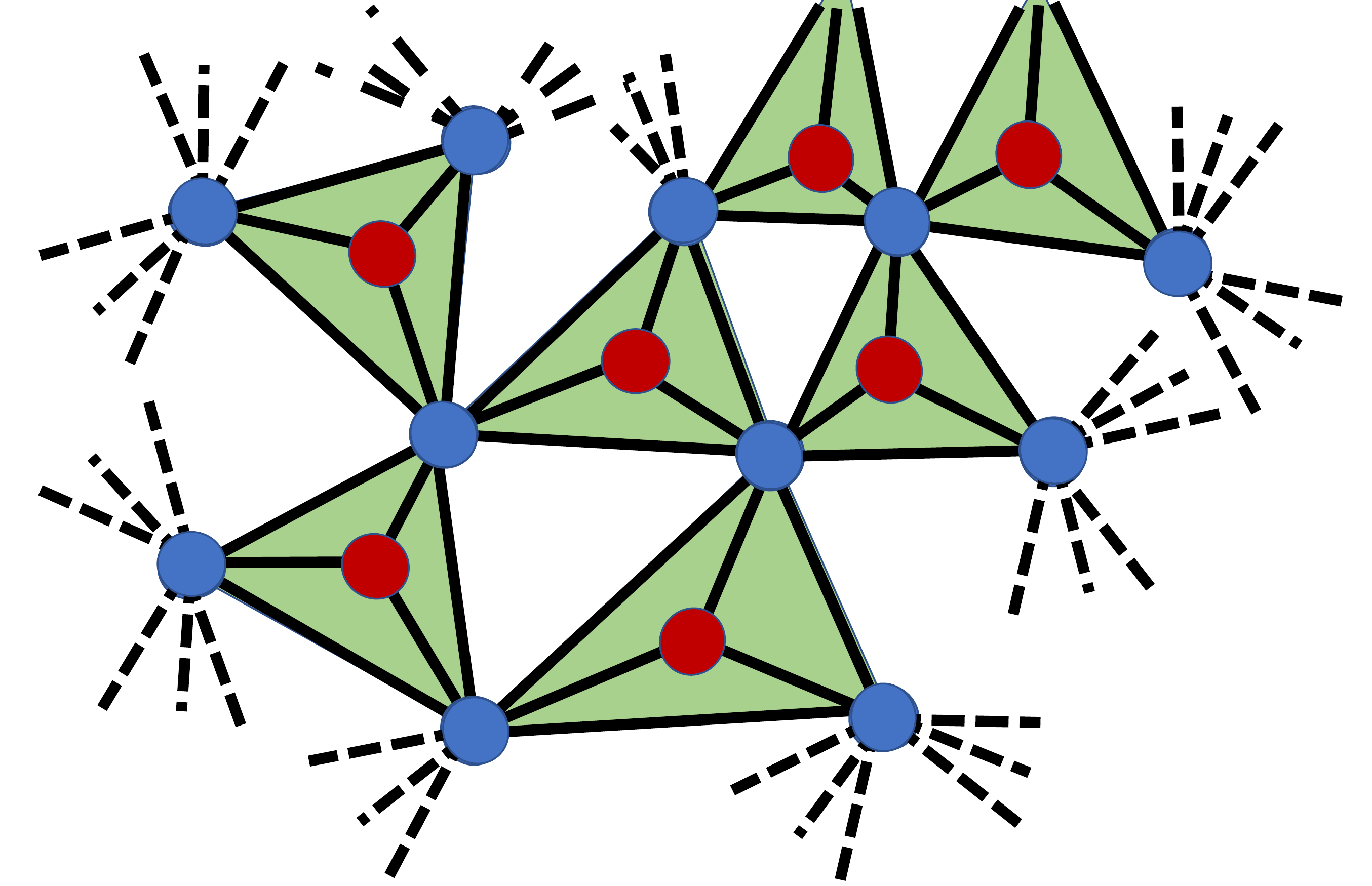}
    \caption{The connectivity graph of a 3-regular 3-XORSAT instance. A 3-XORSAT clause (green) consists of 3 bits (blue) plus an additional auxiliary bit (red), which is used to reduce the clause locality from three-body to two-body. Each bit (or spin) appears in exactly 3 clauses.}
    \label{fig:3r3x_graph}
\end{figure}

\emph{Problem Description.---}
Our benchmark problem instances are derived from $n/2$ randomly generated linear systems of equations of $n/2$ binary variables.  Each equation involves 3 randomly chosen variables, with the constraint that a variable appears exactly in 3 equations. This set of equations can be solved in polynomial time via Gaussian elimination (or conjugate gradient methods~\cite{conjGrad}), so not only do we know the solution to the linear system but we can also ensure that there is a unique solution. The linear system is then mapped to a cubic unconstrained binary optimization instance (specifically a 3-regular 3-XORSAT instance), where each equation in the linear system corresponds to a single 3-XORSAT clause. This class of problems is known to stymie heuristic annealing-type algorithms~\cite{Fra2001,Bar2002,Jor2010,Ric2010,Gui2011,Far2012}, even though the linear system can be solved efficiently. The solution to the linear system is also the solution to the optimization problem after the mapping. Finally, each instance is reduced to a QUBO~\cite{Hen2019}, which inevitably introduces an overhead~\cite{Per2019,Val2020}. Indeed, this final step requires the addition of $n/2$ ancillary bits (one for each clause), forming an $n$-bit QUBO problem. The reduction, which is depicted in Fig.~\ref{fig:3r3x_graph}, preserves the hardness of 3-regular 3-XORSAT, as was demonstrated in Ref.~\cite{Hen2019}. 
We refer to this class of problem instances as the 3R3X planted-solution instances. For a careful statistical mechanical analysis of this class see Bernaschi \textit{et al.}~\cite{bernaschi2021leading} and references therein. Their salient feature is a 
`golf course' energy landscape, characterized by a random first order transition in the language of spin glasses. There exists an exponential (in $n$) number of
local minima below the dynamical critical temperature ($T_d>0$), such
that for temperature $T<T_d$ ergodicity is broken and the equilibration timescale scales exponentially in $n$. In particular, this means that heuristic optimization algorithms based on random bit (spin) flipping -- whether single or as clusters -- are expected to take a time scaling exponentially in $n$ to find the global minimum (ground state).

\emph{Metric Description.---}
Our objective is to quantify how the computational time to find the optimum solution scales with problem size for our suite of specialized solvers.
Since the solvers we study are stochastic, we use the optimal time-to-solution (TTS) metric~\cite{Ron2014,Alb2018}.  For an ensemble of instances of size $n$ drawn from the 3R3X problem class, the optimal TTS (optTTS) for a quantile $q$ is the minimum time required to find the solution at least once with a desired probability of $0.99$ for that quantile:
\beq
\langle \operatorname{TTS} \rangle_q = \min_{t_f} \left\langle t_{f} \frac{\ln \left(1-0.99\right)}{\ln \left[1-p_i\left(t_{f}\right)\right]} \right\rangle_q \frac{1}{f_p(n)}.
\label{eq:TTS}
\eeq
Here $t_f$ is the time the algorithm is run for, and the parallelization factor $f_p(n)$
counts how many replicas of the same instance can be run in parallel on a given solver implementation.
For example, for a solver that accommodates a maximum of $n_{\max}$ bits (or spins), we have $f_p(n)= \lfloor n_{\max}/n \rfloor$. Each replica $r$ (where $r\in \{1,\dots , f_p(n)\}$) results in a guess $E^{(r)}_{i,j}$ of the minimum (or ground state energy) of the cost function~\eqref{eq:cost} for the $j$'th run of the $i$'th instance, and we can take $E_{i,j} = \min_r E^{(r)}_{i,j}$ as the proposed solution for that run. We assume the replicas are independent and that each has equal probability $p_i(t_f)$ of finding the optimal solution. Then, using $f_p(n)$ replicas the probability that at least one will find the ground state is $p'_i(t_f) = 1-(1-p_i(t_f))^{f_p(n)}$, which is also the probability that $E_{i,j}$ is the ground state energy for the $j$'th run of the $i$'th instance. Let $R$ be the number of repetitions required to find the ground state at least once with probability $0.99$; then $0.99 = 1-(1-p'_i(t_f))^R$ and $\text{TTS}=t_f R$, which gives Eq.~\eqref{eq:TTS}, where 
$\langle \cdot \rangle_q$ denotes taking the $q$'th quantile over the distribution of instances~\cite{Alb2018}. 

The TTS metric balances the trade-off between a single long run of the stochastic, heuristic algorithm with a high success probability and multiple short runs with low success probability. Underlying Eq.~\eqref{eq:TTS} is the assumption that different solver runs $j$ are independent with identical success probability for the same instance. 
While in principle the number of runs $R = \frac{\ln(1-0.99)}{\ln(1-p'_i(t_f))}$ should be rounded up to the nearest whole number (at least one run is necessary), we do not apply rounding since this can result in sharp changes in the TTS, which can complicate extracting the scaling behavior~\cite[App.~A]{Alb2018}. We remark that it is important not to confuse the parallelization factor $f_p(n)$ with parallel execution on multiple CPU cores. 
Such parallelization can make the TTS as defined in Eq.~\eqref{eq:TTS} arbitrarily small and also incurs an additional energy/monetary cost that is proportional to the number of CPUs. Rather, as explained above, the parallelization factor applies when a single processor can accommodate multiple replicas.

Using the TTS metric and following the procedure outlined in Methods, we  calculate different quantiles of the TTS for a given problem size $n$ for a given solver.  We optimize the parameters of the solver to minimize this TTS for each size (in particular $t_f$), hence identifying the optTTS. This optimization is essential for avoiding the trap of feigned scaling~\cite{Ron2014}, and without it an extracted scaling exponent can only be trusted as a lower bound on the true value~\cite{Hen2015}. For sufficiently large $n$, we then extract the scaling of optTTS as a function of $n$. 

Finally, we remark that we chose to focus on the optTTS as defined in Eq.~\eqref{eq:TTS} since it has become the standard metric in recent years in efforts to benchmark quantum annealers against other heuristic solvers. Other metrics or methods are certainly possible and may have certain advantages; e.g., a metric based on optimal stopping theory that balances TTS with energy or monetary cost~\cite{Vinci:2016tg}, or an alternative method for computing the TTS that was proposed and used for the same problem in Ref.~\cite{bernaschi2021leading}, and argued to be less computationally demanding (see Methods). 
\begin{center}
\begin{table*}[t]
\bgroup
\def\arraystretch{1.1}
\begin{tabular}{|c||c|c|c|c|c|} \hline
Solver & \begin{tabular}[c]{@{}c@{}}Parallel\\ Tempering\end{tabular} & \begin{tabular}[c]{@{}c@{}}Fujitsu Digital \\Annealer Unit\end{tabular} & \begin{tabular}[c]{@{}c@{}}Toshiba Simulated\\Bifurcation Machine\end{tabular} & \begin{tabular}[c]{@{}c@{}}D-Wave \\
Advantage 1.1\end{tabular} & \begin{tabular}[c]{@{}c@{}}SATonGPU \end{tabular} \\ \hline
\begin{tabular}[c]{@{}c@{}}Hardware\end{tabular} & Single CPU & ASIC & Single GPU &  \begin{tabular}[c]{@{}c@{}}Superconducting\\Qubits \end{tabular} & \begin{tabular}[c]{@{}c@{}} Single GPU\end{tabular} \\ \hline
Connectivity & Full, Dense & Full, Dense & \begin{tabular}[c]{@{}c@{}}small $n$: Full, Dense\\large $n$ Full, Sparse\end{tabular} & Pegasus (Deg. 15) & Full, Dense \\ \hline
\begin{tabular}[c]{@{}c@{}}Max QUBO\\Size $n$\end{tabular} & \begin{tabular}[c]{@{}c@{}}RAM-limited,\\\textgreater{}10,000\end{tabular} & \begin{tabular}[c]{@{}c@{}}8192/4096/2048\\ ~*precision dependent\end{tabular} & \begin{tabular}[c]{@{}c@{}}10,000\\max $10^{6}$ $J_{ij} \neq 0$\end{tabular} &  \begin{tabular}[c]{@{}c@{}}Clique: 128\\3R3X: $\approx$256\\Native: 
5436\end{tabular} & \begin{tabular}[c]{@{}c@{}} RAM-limited,\\\textgreater{}10,000 \end{tabular}\\ \hline
\begin{tabular}[c]{@{}c@{}}Precision \end{tabular} & \begin{tabular}[c]{@{}c@{}}64 bit float\\$\approx10^{-16}$\end{tabular} & \begin{tabular}[c]{@{}c@{}}16/32/64 bit\\(signed int)\\$\approx10^{-4}/10^{-9}/10^{-19}$\end{tabular} & \begin{tabular}[c]{@{}c@{}}64 bit float\\$\approx10^{-16}$\end{tabular} &  \begin{tabular}[c]{@{}c@{}}Noise-limited.\\$\approx$5 bit or $10^{-2}$\end{tabular} & \begin{tabular}[c]{@{}c@{}}64 bit float\\$\approx10^{-16}$\end{tabular} \\ \hline
Parallelization & 1 per CPU core & 8 per DA & \begin{tabular}[c]{@{}c@{}}40 per GPU\end{tabular} & \begin{tabular}[c]{@{}c@{}} $\lfloor n_{\max}/n \rfloor$ replicas \\~ *connectivity dependent \end{tabular} & \begin{tabular}[c]{@{}c@{}} 327680 replicas  \end{tabular} \\ \hline
Accessed Via & USC-UNM code & \begin{tabular}[c]{@{}c@{}} DA Center Japan \\  4/25/20 \end{tabular} & \begin{tabular}[c]{@{}c@{}} Amazon Web \\ Services~\cite{sbm_aws} 8/20/20 \end{tabular} & \begin{tabular}[c]{@{}c@{}} LEAP cloud \\ 10/31/2020 \end{tabular} & \begin{tabular}[c]{@{}c@{}} n/a \end{tabular}\\ \hline
\end{tabular}
\egroup
\caption{Properties of the solvers used in this study, with the exception of MemComputing.
Note that the couplings $Q_{ij}$ in this study were all integer-valued; hence precision does not play a role for the digital devices but still plays a role for the analog one (D-Wave). The first four solvers were all benchmarked by the authors; the SATonGPU solver was benchmarked by Bernaschi \textit{et al.}~\cite{bernaschi2021leading}.}
\label{tab:props}
\end{table*}
\end{center}

\emph{Solver Descriptions.---} We now provide brief overviews of each benchmarked solver, with data collection, detailed algorithm descriptions, and parameter settings elaborated on in the Methods section. Table~\ref{tab:props} summarizes important properties of the solvers we benchmarked. The SATonGPU algorithm and Memcomputing solvers were benchmarked by others, as described below.

\emph{Parallel Tempering (PT):}
Parallel Tempering is a Markov chain Monte Carlo (MCMC) method~\cite{Swe1986,Gey1991,Huk1996} that uses multiple replicas to speed up the equilibration dynamics. Each replica is described by a temperature and a bit/spin configuration, and each replica evolves independently using single-bit/spin Monte Carlo updates according to the Metropolis criterion~\cite{Met1953,Has1970} followed by the exchange of configurations (or equivalently temperatures) between the replicas.

Our implementation of the PT algorithm uses a single CPU core, and we use it to establish a baseline to compare the other QUBO solvers against.  In practice, independent PT simulations can be run on different CPU cores to estimate the success probability of the algorithm more expediently, but we report the TTS of the algorithm using a single core.  In order to predict the TTS of PT on a multi-core processor using our reported data, we need only divide the TTS reported by the number of CPU cores used.
We note that further performance enhancements of PT are possible, e.g. by tuning the temperature distribution~\cite{Kat2006,Zhu2020} and recycling random numbers~\cite{Yat2013}, although we did not implement such enhancements. The Houdayer~\cite{Hou2001} (or iso-energetic cluster (ICM)~\cite{Zhu_ICM_2015}) multi-bit/spin update was tested but found not to provide a performance benefit on our 3R3X instances~\cite{Hen2019}, so we did not use this type of update here.

\emph{Fujitsu Digital Annealer Unit (DAU):}
Fujitsu Limited has built application-specific integrated circuits (ASICs) for performing a parallel tempering type algorithm on QUBO problems called the Digital Annealer~\cite{Matsubara2017,Mat2020,Tsukamoto2017}. The current generation of the DAU allows instances with up to 8192 fully connected variables to be programmed.  There is a limited form of independent parallelization on the device, as it can be programmed to run 8 simultaneous replicas of 1024 fully connected variables. We use the second mode, giving the device we benchmarked a parallelization factor $f_p=8$ for all our instances.

The DAU implements `rejection-free' Monte Carlo updates~\cite{RFMC} and dynamical temperature adjustment, both of which are important algorithmic differences from our parallel tempering implementation. See Algorithm \ref{algo:DAU} in Methods for details.  

\emph{Toshiba Simulated Bifurcation Machine (SBM):} The SBM~\cite{Got2019b} uses the classical dynamics of Kerr-nonlinear parametric oscillators~\cite{Got2019} to solve QUBO/Ising problems.
We benchmarked the publicly available numerical implementation of the SBM on AWS Marketplace~\cite{sbm_aws}, which uses 8 Nvidia V100 Tensor Core GPUs.

\emph{D-Wave Advantage:}
The D-Wave Advantage (DWA) implements the quantum annealing algorithm~\cite{Kad1998,Brooke1999,brooke_tunable_2001,Far2001,Kaminsky-Lloyd} in programmable hardware; it uses arrays of superconducting Josephson junctions to emulate the low energy spectrum of a time-dependent transverse field Ising Hamiltonian~\cite{Dwave,Ber2010,Har2010,Bun2014}. It is the only solver in our study that does not natively implement all-to-all connectivity. A minor-embedding procedure~\cite{Choi1,Choi2} is necessary to fit the 3R3X instances onto the Pegasus connectivity graph of the DWA~\cite{Boo2020}, increasing the physical number of variables used to implement the problem instances beyond $n$. See Methods for more details.

\emph{Virtual MemComputing Machine (MEM):}
Our 3R3X instances were run on a Virtual MemComputing Machine (VMM) by a MemComputing Inc. team member, and we used the data to perform our optTTS benchmark. The VMM uses classical equations of motion to simulate the theorized Digital MemComputing Machine~\cite{Tra2015,Tra2018,Div2018}. 

\emph{SATonGPU:}
The `golf-course' structure of the optimization landscape, meaning that problem hardness is mainly due to entropic barriers, is well suited to a quasi-greedy algorithm~\cite{bellitti2021entropic} that is not allowed to jump over large energy barriers and does not satisfy detailed balance. Instead, a quasi-greedy algorithm performs with high probability an energy-decreasing step when possible, but when a local minimum is reached a bit/spin participating in at least one violated interaction is flipped. The algorithm chooses a random variable at each time step and flips it with a probability that depends on how many of the clauses associated with that variable are satisfied. In particular, a spin involved only in satisfied interactions is not flipped, thus ensuring that once the ground state is found, the algorithm terminates. Such an algorithm was implemented by Bernaschi \textit{et al.}.~\cite{bernaschi2021leading}, who in addition took advantage of massive parallelism of GPUs to run, instead of a single instance evolving for a very long time, a large number $f_p(n) =  327680$ of independently evolving replicas for a short time, each one starting from a different random initial condition (see Methods for a more complete description). Note that this `SATonGPU' algorithm optimizes the original \emph{cubic} version of the 3R3X instances, not the QUBO version that all our other solvers were given.

\section*{Results}
\begin{figure*}[thbp]
    \centering
    \includegraphics[width=16cm]{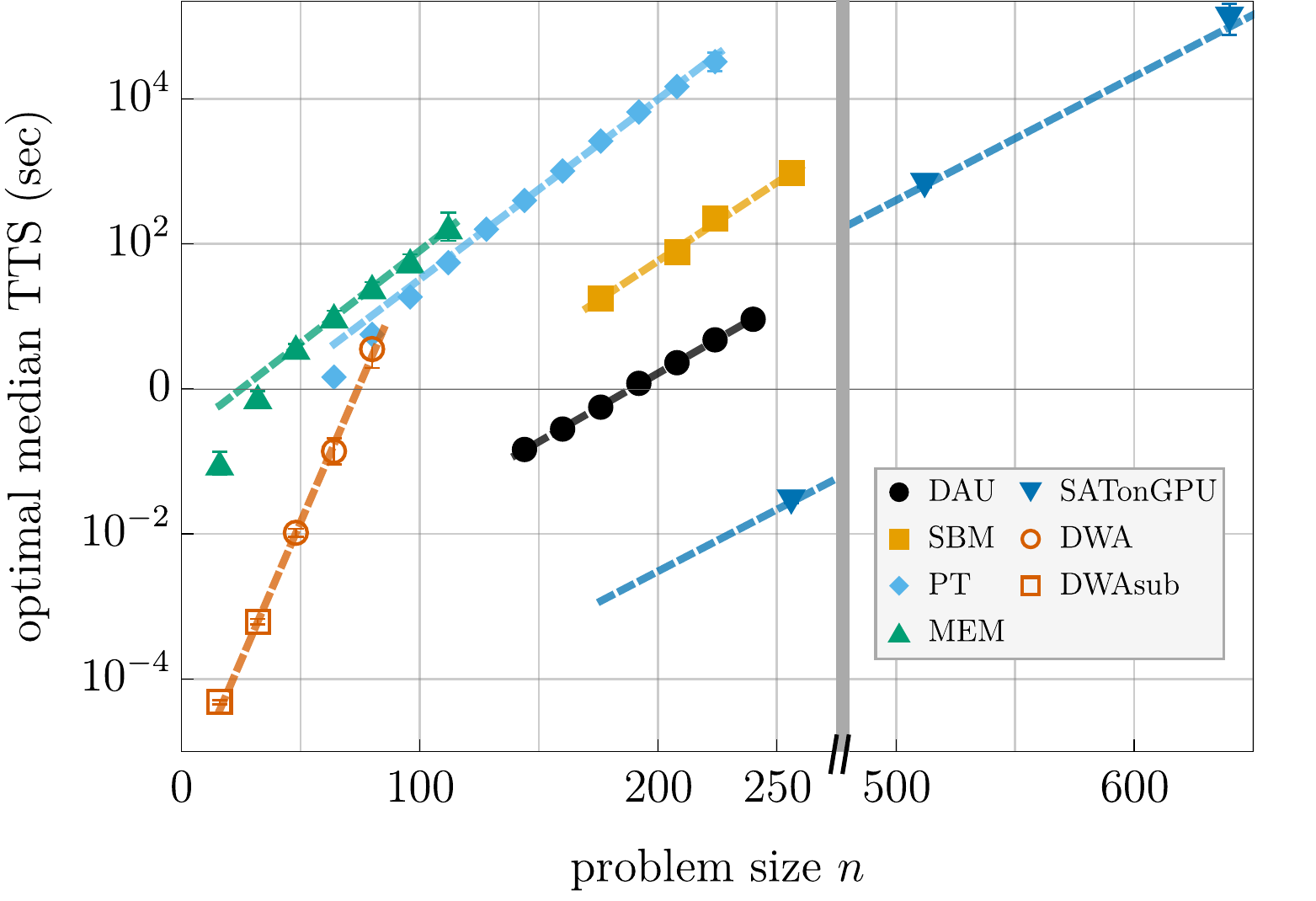}
    \caption{Median optTTS for different solvers for the quadratic 3R3X instances at different problem sizes $n$.  
    The error bars correspond to $2\sigma$ confidence intervals calculated using a Bayesian-bootstrap. Each solver is represented with a different marker and color, as denoted by the legend. DAU is Fujitsu's Digital Annealer Unit, run in parallel mode on 25 April 2020. SBM is Toshiba's Simulated Bifurcation Machine, accessed via Amazon Web Services on 20 August 2020. PT is our implementation of parallel tempering on a single CPU core. MEM is the Virtual MemComputing Machine. SATonGPU is the data from Fig.~5 of Bernaschi \textit{et al.}~\cite{bernaschi2021leading}, after converting their native three-body 3R3X results in $n/2$ variables to $n$ two-body variables by simplying doubling their reported $n$ values. Note that the SATonGPU results are plotted both in the left panel (ending around $n=250$) and the right panel (starting around $n=500$), as this solver reached significantly larger problem sizes than the other solvers; its optTTS is smaller by at least two orders of magnitude than the rest. DWA is the D-Wave Advantage1.1 device accessed via LEAP on 31 October 2020. DWAsub are suboptimal points in which the optimal runtime is below $1\mu s$, the lowest runtime possible on the Advantage1.1 device. 
     The lines correspond to the exponential fits [Eq.~\eqref{eq:fit}] of the data, with the coefficients given in Table~\ref{tab:aB_values}. The DWAsub points were not used in computing the DWA scaling exponent reported in Table~\ref{tab:aB_values}.}
    \label{fig:3R3Xmain}
\end{figure*}
We fit each solver's 
optTTS as a function of 3R3X instance size $n$ to the model 
\beq 
\label{eq:fit}
\langle \text{TTS} \rangle_{q} \sim   10^{\alpha n+\beta} \ ,
\eeq
with the quantile $q\in\{0.25,0.5,0.75\}$.
We find that this two-parameter model is sufficient to fit the data from each solver. We do not have enough data points to obtain a reliable fit to a more elaborate model such as $10^{\alpha n+\beta + \gamma\log n}$.

We show our main result in Fig.~\ref{fig:3R3Xmain}, where we plot the median ($q=0.5$) optTTS of all the solvers we have benchmarked, along with the SATonGPU and MEM solvers. 
The scaling exponent $\alpha$ and the prefactor $\beta$ for each solver are summarized in Table~\ref{tab:aB_values}. 

\begin{table*}
\centering
\begin{tabular}{|c|c|c|c|c|c|c|} 
\hline
         & \multicolumn{3}{c|}{$\alpha$}                                   & \multicolumn{3}{c|}{$\beta$}                                     \\ 
\hline
Solver   & $q=.25$  & $q=.5$                                    & $q=.75$  & $q=.25$  & $q=.5$                                    & $q=.75$   \\ 
\hline
SATonGPU & n/a      & {\cellcolor[rgb]{0.82,0.82,0.82}}.0171(7) & n/a      & n/a      & {\cellcolor[rgb]{0.82,0.82,0.82}}-5.9(3)  & n/a       \\ 
\hline
DAU      & .0181(2) & {\cellcolor[rgb]{0.82,0.82,0.82}}.0185(4) & .0190(4) & -3.51(4) & {\cellcolor[rgb]{0.82,0.82,0.82}}-3.56(7) & -3.49(7)  \\ 
\hline
SBM      & .0211(7) & {\cellcolor[rgb]{0.82,0.82,0.82}}.0217(6) & .0234(8) & -2.6(2)  & {\cellcolor[rgb]{0.82,0.82,0.82}}-2.6(1)  & -2.7(1)   \\ 
\hline
PT       & .0239(1) & {\cellcolor[rgb]{0.82,0.82,0.82}}.0248(2) & .0252(1) & -0.92(2) & {\cellcolor[rgb]{0.82,0.82,0.82}}-0.97(4) & -0.97(2)  \\ 
\hline
MEM      & .030(9)   & {\cellcolor[rgb]{0.82,0.82,0.82}}.025(2)  & .024(3)  & -1(2)    & {\cellcolor[rgb]{0.82,0.82,0.82}}-0.6(2)  & -0.2(2)   \\ 
\hline
DWA      & n/a    & {\cellcolor[rgb]{0.82,0.82,0.82}}.08(4)   & n/a   & n/a    & {\cellcolor[rgb]{0.82,0.82,0.82}}-6(2)    & n/a     \\
\hline
\end{tabular}
\caption{Fit parameters $\alpha$ and $\beta$ of Eq.~\eqref{eq:fit} for the first quartile, median, and third quartile optTTS data corresponding to Fig.~\ref{fig:3R3Xmain} (which shows the median only). The uncertainty on the last significant digit is indicated in parameters and reflects $2\sigma$ confidence intervals. Results are sorted by lowest $\alpha$ for $q=0.5$. In all cases $\alpha$ was extracted after optimization of the TTS (see text), so that it reflects the true value of the scaling exponent. We did not have enough data to report the first and third quartiles of SATonGPU and DWA.
}
\label{tab:aB_values}
\end{table*}

As is clear from both Fig.~\ref{fig:3R3Xmain} and Table~\ref{tab:aB_values}, the solvers separate into three groups by the scaling exponent $\alpha$: (i) SATonGPU and DAU, (ii) SBM, PT and MEM, (iii) DWA. Between the two leading solvers in group (i), the SATonGPU solver has both a slightly smaller scaling exponent and by far the best absolute time to solution (see the $\beta$ parameter in Table~\ref{tab:aB_values}). The latter is attributable to the massive parallelization advantage enjoyed by this solver, allowing it to run 327680 replicas (see Table~\ref{tab:props}). DAU, despite running the QUBO version of the 3R3X problem, achieves nearly the same scaling as SATonGPU, but the two are separated to within $2\sigma$ statistical confidence. Notably, DAU has an advantage in both scaling and absolute time over our PT implementation, indicating that both hardware and algorithmic optimization can make a substantial difference. For example, by choosing a temperature distribution that has a constant swap probability of 0.23 between neighboring temperatures, the PT simulations of Ref.~\cite{bellitti2021entropic} using the original cubic form of 3R3X has a scaling that approaches that of the quasi-greedy algorithm.
It is also worth noting that, as seen from Table~\ref{tab:aB_values}, the group (ii) solvers are separated to within $2\sigma$ statistical confidence from the group (i) solvers. 
Finally, it is clear that DWA has the highest scaling exponent of the group, and is likewise separated from both groups (i) and (ii).

\section*{Discussion}
We have benchmarked QUBO solvers implementing various algorithmic approaches on one particular class of problem instances. The problem class is given by a mapping of linear systems of equations representing 3-regular 3-XORSAT combinatorial optimization problems in $n/2$ variables to QUBO instances in $n$ variables, and exhibits the characteristic exponential scaling with $n$ of the solution time of NP-hard problems, due to its golf-course-like energy landscape. Furthermore, these are planted-solution problems, with the property that the solution of the linear system (which is easily found) is also the solution of the QUBO problem. The heuristic solvers we benchmarked only have access to the QUBO form of the problem (or the original 3-XORSAT form in the case of SATonGPU), not the original linear system. This provides a useful class of problems against which to benchmark QUBO solvers, but we note that this problem class may not be representative of the performance of these solvers on other classes of problems. For example, the instances do not exhibit an all-to-all connectivity graph, and their precision requirements are minimal (the QUBO parameters were integers and fell in the range of [-30,16]).

Only two of the solvers studied here represent dedicated hardware: DWA and DAU. The rest (PT, SBM, MEM, and SATonGPU) were benchmarked on standard digital classical hardware (CPUs or GPUs; see Table~\ref{tab:props}). However, both SBM and MemComputing have been proposed as approaches that can also be implemented in dedicated hardware, and we might expect some performance improvement to accompany such an implementation.  
We remark that due to lack of access, our study leaves out the Coherent Ising Machine, which is another promising solver based on dedicated hardware~\cite{McM2016,Yam2017,ng2021efficient}.

The DWA exhibits the highest optTTS scaling in our study. Importantly, it is the only analog device among the ones we benchmarked; for all the other solvers the QUBO problem can be specified digitally within their precision range. The median minimum gap along the interpolating Hamiltonian of the 3-XORSAT instances (in its native three-body form and before minor-embedding) closes exponentially as $10^{-0.0174n}$~\cite{Far2012}.  If we assume that the adiabatic runtime scales as the inverse-gap squared (although a rigorous worst-case bound predicts an inverse-gap cubed scaling~\cite{Jan2007}), then this yields a runtime scaling exponent of $\alpha=0.0348$ for the adiabatic algorithm. 
Comparing this to the $\alpha$ values given in Table~\ref{tab:aB_values}, it is lower than the observed DWA value. While we do not know how the mapping to the QUBO form affects the gap scaling, a difference in scaling between the adiabatic algorithm and DWA would not be surprising since the DWA does not provide a direct implementation of the adiabatic algorithm for a variety of reasons, all of which are expected to lead to reduced performance. First and foremost, it does not operate as a closed quantum system, and thermal excitation and relaxation effects usually play a detrimental role in its performance as an optimizer or sampler~\cite{Amin:2015qf,Boixo:2014yu,Alb2018,Albash:2017ab} (with a few exceptions, e.g., Ref.~\cite{DWave-16q}). In addition, we expect that minor-embedding and other sources of noise such as cross-talk~\cite{q-sig2} and integrated control errors (ICE)~\cite{dwave-manual} further hurt performance~\cite{Zhu2016,Alb2019}. While error suppression and correction can act to partly mitigate these adverse effects~\cite{Pearson:2019aa}, the baseline of a large $\alpha$ value even in the ideal case of a closed quantum system suggests that quantum annealing is at a disadvantage in competing with state of the art classical approaches such as the ones we have explored here.  However, it should be noted that the median minimum gap of the instances occurs consistently at the midpoint of the interpolating Hamiltonian, suggesting that a local adiabatic schedule could help reduce the runtime dependence to just the inverse gap~\cite{Roland2002} (i.e., $\alpha=0.0174$), making it on-par with the best scaling observed in this study, as long as we ignore the $O(\log(n))$ due to the numerator of the adiabatic condition~\cite{Alb2018b}; such a correction was also ignored in our optTTS fits for all other solvers, as mentioned above.
 Alternative annealing protocols could also overcome this problem and might yet prove to be beneficial~\cite{crosson2020prospects}, but are beyond the scope of the present study. 

We found the Fujitsu DAU to have the lowest optTTS and scaling exponent among the solvers that solved the QUBO form of the 3R3X instances. The SATonGPU algorithm is statistically distinguishable to within $2\sigma$ confidence from the DAU in terms of optTTS scaling, and has a slightly smaller scaling exponent. The Toshiba SBM scales better than our base implementation of parallel tempering. 

The SATonGPU algorithm has by far the lowest optTTS, as seen in Fig.~\ref{fig:3R3Xmain}. Recall that it solved the native three-body form of the instances, unlike all our other solvers.
While locality reduction from three-variable to two-variable clauses incurs an overhead~\cite{Val2020}, and it is unclear whether avoiding this reduction provided SATonGPU a scaling exponent improvement over solving the QUBO form, what is clear is that the massive parallelization achieved by the GPU-optimized algorithm provides approximately $5$ orders of magnitude reduction in the TTS. One might be tempted to interpret this as an indication of the futility of employing dedicated hardware, but we believe it instead indicates the limitations of using a single problem class for general benchmarking; in this case, the relatively simple structure of the problem class allowed for the implementation of a highly optimized algorithm. Nonetheless, we hope that given its attractive properties, the 3R3X problem class will remain a useful standard for the benchmarking of both existing and future QUBO/Ising solvers and specialized devices.

Given the limited generalizability of conclusions drawn based on a single instance class, future work will employ a varied set of problem classes for benchmarking across different general-purpose solvers. We expect that different solvers will be better suited for different problem classes, and a fruitful approach will be to try a suite of solvers in attacking a given NP-hard optimization problem.% rather than trying to identify a single winner. 

Another important future direction that our work does not address since the 3R3X instances are not well suited to it is that of \emph{approximate} optimization. There too we expect different solvers to excel on different problem classes.

\section*{Methods}
\subsection*{Bayesian bootstrap for determining optimal time to solution}
In order to calculate optTTS [Eq.~\eqref{eq:TTS}], we need to estimate $p_i(t_f)$, the success probability of one run of a particular instance $i$ for a given runtime $t_f$ of the algorithm. Instead of using a sample of runs as a point estimate for $p_i(t_f)$, we use Bayesian inference to estimate the binomial distribution of $p_i(t_f)$ for each problem instance as in Ref.~\cite{Hen2015}. Our prior is a beta distribution $\text{Beta}(\alpha,\beta)$ with $\alpha=\beta=0.5$ (the Jeffery's prior), chosen because it is invariant to reparameterizations of the space and learns the most~\cite{Job:2017aa,Job2018}. The resulting posterior distribution for $p_i(t_f)$ of a particular instance is then $\text{Beta}(0.5+n_S,0.5+N-n_S)$, where $N$ is the number of runs and $n_S$ the number of successful runs. 

To estimate optTTS for all $I$ instances (we have 100 instances for each problem size $n$) of a given problem size, we pick a regular distribution of $t_f$'s, and we find the posterior distributions $\pi(p_i(t_f))$ for each instance $i \in \{1,2,...,I\}$. For each instance, we can then calculate the instance TTS for the runtime $t_f$:
\beq
\text{TTS}_i(t_f) = t_f \frac{\ln[1-0.99]}{\ln[1- p_i(t_f)]} \ . \frac{1}{f_p}
\eeq
We then perform a bootstrap according to Algorithm \ref{algo:TTSboot} to determine the average of a quantile $q$ of the TTS, $\langle \text{TTS}_i(t_f) \rangle_q$. For each quantile $q$, we then define optTTS as the minimum value of the TTS attained as a function of $t_f$ [as defined by Eq.~\eqref{eq:TTS}], which identifies a corresponding optimal value of the runtime $t_f$.
Our definition of optTTS is over the entire set of instances of a given size and for a particular quantile and not for a single instance. If the minimum TTS is such that $\frac{\ln[1-0.99]}{\ln[1- p_i(t_f)]} < 1$, then we set the optimal runtime $t_f^\ast$ such that $\frac{\ln[1-0.99]}{\ln[1- p_i(t_f^\ast)]} = 1$.  This corresponds to only needing to run the algorithm once.
\begin{algorithm}[H]
    \caption{Bayesian Bootstrap for $\langle \text{TTS}_i(t_f) \rangle_q$}
  \label{algo:TTSboot}
   \begin{algorithmic}[1]
   \State Choose the number of bootstrap resamples $R$ ($R=1000$)
   \For{$r=1,2,...R$}
   \State sample $I$ instances with replacement.  Denote this set of instances by $\mathcal{I}_r$.
   \State Find the $q$'th quantile of the set $\{\text{TTS}_{i}(t_f)\}_{i \in \mathcal{I}_r}$ and denote it by $\langle \text{TTS}_r(t_f) \rangle_{q}$
   \EndFor
   \State The mean of the distribution $\left\{\langle\text{TTS}_r(t_f)\rangle_{q}\right\}_{r=1}^R$ is taken as an estimate for $\langle\text{TTS}(t_f)\rangle_q$ for instances from the problem class, with the standard deviation of $\left\{\langle\text{TTS}_r(t_f)\rangle_{q}\right\}_{r=1}^R$ giving the $1\sigma$ confidence interval.
   \end{algorithmic}
\end{algorithm}
While in our discussion above we have only explicitly optimized over the runtime $t_f$, optTTS and subsequent optimal scaling analysis requires all solver parameters to be properly optimized. In our discussion of each solver, we mention the extent to which we have optimized these parameters.

\subsection*{Parallel Tempering (PT)}
A single PT step includes both Monte Carlo (MC) updates and replica exchanges (or replica swaps). The replicas are evolved independently by performing a fixed number of Monte Carlo (MC) `sweeps,' where a single sweep corresponds to attempting to update every variable in every replica  successively according to the Metropolis criterion~\cite{Met1953,Has1970} for calculating the acceptance probability $\mathcal{A}$ of a bit/spin flip:
\beq
\mathcal{A} = \min \{1,e^{-\beta \Delta E}\}\ ,
\eeq
where $\beta$ is the inverse temperature of the replica, and $\Delta E$ is the change in energy associated with the spin flip. 

After the completion of the MC sweeps, we perform a replica swap update.  The spin configuration of all neighboring replicas in temperature are swapped with a probability satisfying detailed balance:
\beq
\mathcal{A} = \min \{1,e^{\Delta \beta \Delta E}\}\ ,
\eeq
where $\Delta \beta$ is the difference in inverse-temperatures of the two replicas being swapped, and $\Delta E$ is the difference in their energies.
(Equivalently, we can hold the replica configurations fixed and swap the temperatures.) 
\begin{algorithm}[H]
  \caption{Parallel Tempering}
  \label{algo:PT}
   \begin{algorithmic}[1]
   \State initialize replicas to random states
   \State set replicas to log-uniformly distributed temperatures
   \For{each MC sweep}
   \For{each replica (in parallel)}
   \For{each variable}
   \State propose a flip according to Metropolis criterion
   \State if accepted, update state and effective fields
   \EndFor
   \EndFor
   \For{each sequential pair of replicas}
    \State propose a replica exchange (Metropolis)
    \State if accepted, swap replica temperatures.
   \EndFor
   \EndFor
   \end{algorithmic}
\end{algorithm}

\emph{Parameter Settings}.---%
We used 32 replicas with a log-uniformly distributed (normalized to the maximum magnitude Ising parameter) inverse-temperatures $\beta$, ranging from $\beta_{\min}=.1$ to $\beta_{\max}=20$.  The temperature distribution was held fixed during the PT simulation.  We used 10 MC sweeps between each replica swap update, as described above. Setting $\beta_{\max}=5$ made no difference in our results.

\emph{Hardware Specifications}.---%
We used Intel Xeon E5-2680 2.4GHz with dual 10 core processors.  We get a typical PT step (32 replicas, with 10 sweeps per replica, and replica swaps between all neighbor pairs) time of $1.7\mu s \times n$.

\emph{Data Collection}.---%
For each of the 100 instance within a problem size $n$, we ran between $10^3$ and $10^4$ independent PT simulations.  For each simulation, we record the first time the ground state is found and terminate the simulation.  Using the independent runs and their stopping times, we can estimate the cumulative distribution function for the success probability as a function of $t_f$ using a Bayesian bootstrap over the runs to give error estimates.  This allows us to then estimate $p_i(t_f)$ for arbitrary $t_f$ values.

\subsection*{Fujitsu Digital Annealer (DAU)}
A prior study by Aramon \textit{et al.}~\cite{Ara2019} benchmarked an early version of the DAU, referring to it as the parallel tempering Digital Annealer (PTDA), using a similar scaling analysis as ours but on different sets of instances. The early PTDA did not have replica exchanges implemented on the ASIC hardware and was performing that step on a CPU, so we expect better performance in the newer version of the DAU used in our study, where replica exchange is performed in dedicated hardware. 

The DAU algorithm features several differences from the PT algorithm  \ref{algo:PT}. A key difference is the implementation of ``jump" or ``rejection-free" Monte Carlo Markov chains. The idea of rejection-free PT is explored in Ref.~\cite{Ros2020}. The key idea is to avoid the
inefficiency of rejections by exploiting parallelism to compute all
potential acceptance probabilities at once and accept an update move every time. Care must be taken to prevent biased estimates due to convergence to the wrong distribution, as would be the case with uniform sampling; Instead, the `jump chain' is used~\cite{Ros2020}. 

\begin{algorithm}[H]
  \caption{Digital Annealer U2 (DAU)}
  \label{algo:DAU}
  \begin{algorithmic}[1]
   \State initialize replicas to random states
   \State set replicas to log-uniformly distributed temperatures
   \For{each MC step (iteration)}
   \For{each replica, in parallel}
   \For{each variable, in parallel}
   \State propose a flip using $\Delta E_{j} - E_{\text{offset}}$
   \State if accepted, record
   \EndFor
   \If{at least one flip accepted}
   \State choose one flip uniformly at random
   \State update states and effective fields, in parallel
   \State $E_{\text{offset}} \leftarrow 0$
   \Else 
   \State $E_{\text{offset}} \leftarrow E_{\text{offset}} + \text{offset}\_\text{increment}$
   \EndIf
   \EndFor
   \If{due for replica exchange}
   \State update temperatures
   \For{each sequential pair of replicas}
    \State propose a replica exchange (Metropolis)
    \State if accepted, swap replica temperatures.
   \EndFor
   \EndIf
   \EndFor
   \end{algorithmic}
\end{algorithm}

\emph{Algorithm.---}% 
The DAU algorithm \ref{algo:DAU} replaces the serial MC sweep step from PT with a parallel calculation of every Metropolis bit/spin flip update. From the bits/spins that are proposed to be updated according to the Metropolis criterion, a single bit/spin is chosen at random and updated. This process is treated as a single MC step/iteration. With lower bounded probabilities, the process approximates the intended rejection-free roulette-wheel type selection as proven by Liposki \textit{et al.}~\cite{Lip2012}. However, in this implementation of the DAU, it is still possible for every spin in the parallel calculation to be rejected, resulting in repeated MC iterations with no configuration updates. To mitigate this inefficiency, after a failed configuration update, the acceptance probabilities of each variable $\mathcal{A}_j$ are artificially increased by an added energy offset $E_{\text{offset}}$. 
After the machine configuration updates, effective fields are updated in parallel, an $O(n)$ speedup relative to the same step in our parallel tempering implementation.
Temperatures are actively adjusted according to a scheme by Hukushima \textit{et al.}~\cite{Huk1999} to achieve an equal replica-exchange acceptance probability for all adjacent temperatures.

\emph{Parameter Settings.---}%
The ``repex\_interval" parameter for the number of MC steps between replica exchanges was held fixed at $2500$, the result of a coarse optimization. The ``repex\_mode" parameter was set to 2, so the temperature distribution was automatically adjusted during the run without fixed temperature bounds.
The number of replicas was set to the minimal value of 26. The ``offset\_mode" parameter was set to 2 for automatic offset adjustment of the acceptance probabilities. The initial state of the replicas was randomized and a different random seed for the MC updates chosen for each run. Every run was performed with the DAU set at the minimal problem size of 1048 spins, and unused spins/couplings were set to maximally negative values.

\emph{Data Collection.---}% 
The DAU lacks a termination condition, and does not register the time to solution, only the number of MC steps to solution. To estimate $p_i(t_f)$, we repeat instance $i$ for $N$ times ($\approx 200$) with some maximal runtime $t_f^{\max}$ and record the first time $t_0$ the optimal solution is reached. Then we choose a regular distribution of $t_f$'s and update the Jeffery's prior distribution for $p_i(t_f)$ by resampling with replacement from our data. If the run we pull succeeded before (after) $t_f$, it counts as a success (failure), and the usual Baysian Bootstrap for $\langle \text{TTS}_i(t_f) \rangle_q$ from Algorithm \ref{algo:TTSboot} is used.

When attempting to determine the parallelization factor for the DAU, an $n=228$ 3R3X problem was embedded 4 times into a single 1048 spin run of the DAU, with unused spins/couplings set to maximally negative values. The MC updates to solution and runtime was at least four times longer than the average MC updates to solution and runtime of the same $n=228$ problem run alone. As such, there appears to be no parallization benefit within the same 1048 spin section. However, a single DAU has 8192 spins that can be broken into independent 1048 spin machines and operated separately. As such, in our scaling analysis for the DAU, a parallelization factor of $f_p = 8$ is used for every size we studied.

\subsection*{Toshiba Simulated Bifurcation Machine}
The SBM involves solving a series of coupled ordinary differential equations of the form
\beq\label{eq:SBM}
\ddot x_i = -(x^2_i - p(t)+1)x_i + C\left(h_i+\sum_j J_{ij}x_j\right),\quad x_i \in \mathbb{R}
\eeq
via a symplectic Euler method. The parameter $C$ is the relative weight of the Ising parameters $\left\{J_{ij},h_i \right\}$, 
and $p$ is steadily increased from $0$ to $1$ with a time step $dt$. The signs of the final $x_i$ values are taken as $-1$ and $1$ respectively to solve the Ising problem defined by $\left\{J_{ij},h_i \right\}$. The conversion to a QUBO problem is straightforward.
\begin{figure*}[thbp]
    \centering
    \includegraphics[width=16cm]{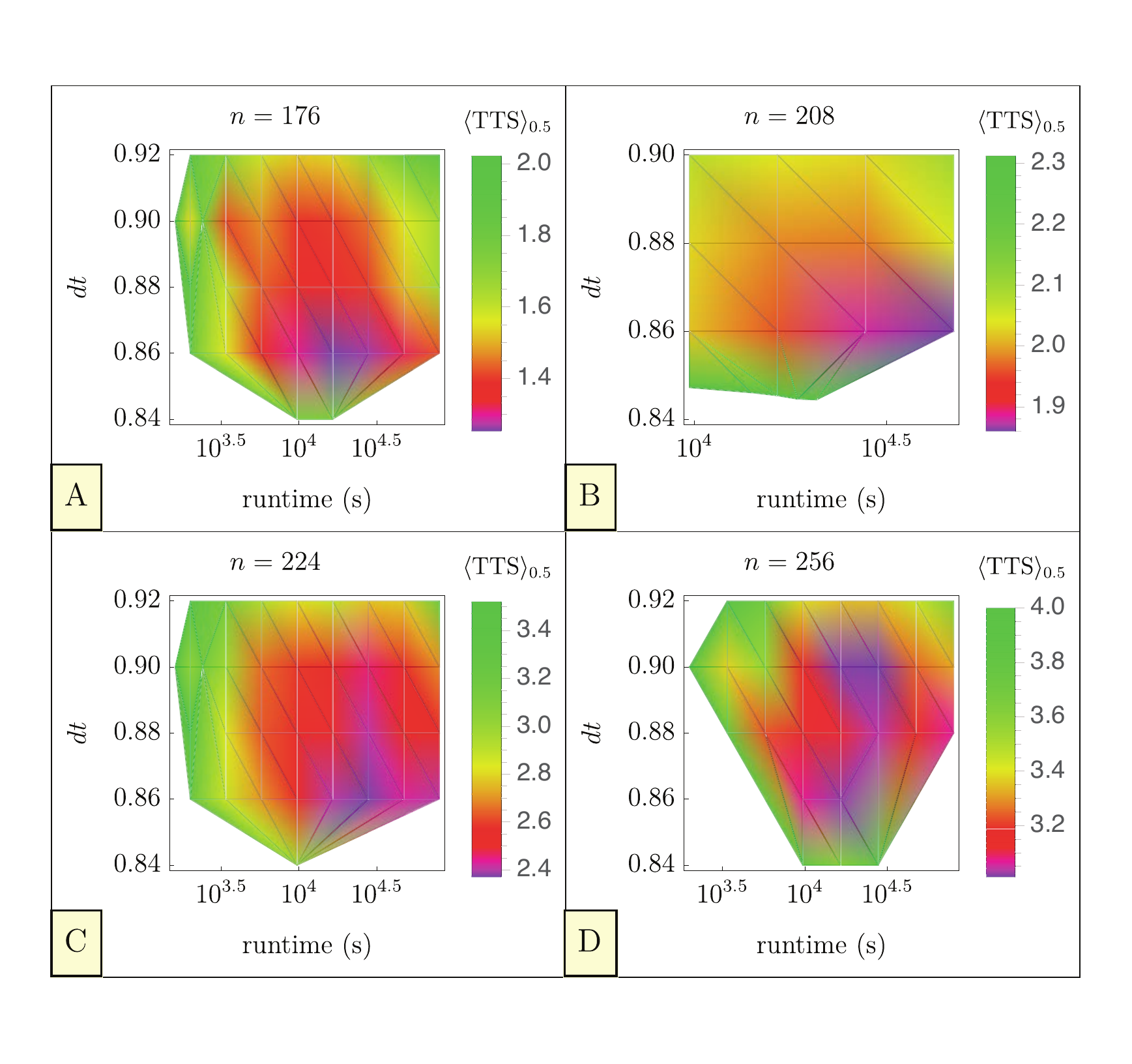}
    \caption{Plots depicting the full $\langle \text{TTS}\rangle_{0.5}$ data collected on the Toshiba Simulated Bifurcation Machine.}
    \label{fig:sbm_opt}
\end{figure*}

\emph{Parameter Settings.---}% 
From Eq.~\eqref{eq:SBM}, there are two parameters beyond the runtime to be optimized: $C$ and $dt$ -- the time step taken by the symplectic Euler numerical solution to the coupled ODE's. We use the auto-adjustment feature provided for $C$, and manually optimize $dt$ around a central value of $.9$ used in previous benchmarks~\cite{Got2019b}. Figure \ref{fig:sbm_opt} depicts the resulting optimization data (without error) used to determine the median optTTS ($\langle \text{TTS}\rangle_{0.5}$) for each value of $n$ in Fig.~\ref{fig:3R3Xmain}.

\emph{Data Collection.---}% 
The 8 Nvidia Tensor Core GPUs run 320 replicas of the algorithm at once. Taking a single Nvidia Tensor GPU for our benchmark, we set $f_p=40$. The returned time $t_{r}$ is precise to only $10^{-2}$ seconds. To resolve anneal times $t_f$ below $10^{-2}$ seconds it is necessary to run $m$ ``loops" of the algorithm and then take $t_r/m$ to determine $t_f$.  Runtime is not directly specifiable; instead one specifies the number of Euler steps to indirectly set runtime. 

We use a log-uniform set of steps from 2000 to 80,000, with an $n$-dependent total time $t_r$ from $100$ seconds to $300$ seconds. The number of loops $m$ is simply the maximum attainable within $t_r$. Each point in the parameter mesh is repeated at least 3 times to average over variation in the auto-adjustment of $C$, and includes an $n$-dependent number of instances from $40$ to $20$. Extra data was collected at points with poor certainty in TTS.

\subsection*{D-Wave}
The D-Wave processor implements a time-dependent Hamiltonian $H(s)$, where $s$ is a dimensionless time parameter increasing from 0 to 1 in the standard `forward' annealing mode we implemented. The Hamiltonian interpolates between the transverse-field contribution and the classical Ising term, reading
\begin{equation}
H(s) = -A(s) \sum_{i} \hat{X}_{i} + B(s) \left(\sum_i h_{i}\hat{Z}_{i} + \sum_{i\neq j} J_{ij}\hat{Z}_i\hat{Z}_j\right), 
\end{equation} 
where $A(s)$ and $B(s)$ time-dependent functions determining the annealing schedule and are fixed by the quantum hardware. $A(s)$ decreases monotonically to 0, while $B(s)$ increases monotonically from near 0.  $\{\hat{X}_i, \hat{Z}_i\}$ are the Pauli operators acting on qubit $i$. At the end of each annealing run ($s=1$) the hardware returns the $\pm 1$ eigenvalues of each $\hat{Z}_{i}$, which are determined by the direction of circulating persistent current in a superconducting metal loop~\cite{Ber2010,Har2010}. Again, the conversion to a QUBO problem is straightforward.

\emph{Data Collection.---}% 
Quantum annealers have a number of unique challenges to overcome in the estimation of $p_i(t_f)$. A embedding is required, introducing additional parameters that require optimization: chain strength, and chain break post-processing. In minor embedding, multiple qubits are chained together via strong ferromagnetic couplings~\cite{Choi1,Choi2,klymko_adiabatic_2012}.  For sufficiently strong ferromagnetic couplings this ensures that the ground state of the Ising Hamiltonian is unaffected by the minor embedding, but it affects the spectrum along the anneal in unpredictable ways, making the strength of these connections critical: too high, and the logical qubit will not flip readily enough to solve the Ising problem, too low and the logical qubit will ``break" as its member physical qubits resolve into different final states that do not align~\cite{marshall2019perils}. With large problems and good tuning, a small number of chains still break, and it is necessary to choose how to map a broken chain to a logical state. At first, we sought to optimize all parameters on a \emph{per size (per $n$)} basis: annealing time, chain strength, and embedding. Using the minorminer feature ``$find\_clique\_embedding()$" that uses heuristics from Ref.~\cite{Boo2016,Zbinden:2020wv} to adapt the optimal embeddings to broken graphs, we found that a per $n$ optimization via full clique embeddings at each $n$ produced immeasurably high TTS after $n>32$, even with the use of ``$VirtualGraphComposite()$'', an advanced method for tuning logical qubits for a particular embedding. (This method involves correcting the flux biases of the physical qubits that are skewed due to extended-range ferromagnetic couplings used to form logical qubits.) Therefore, out of necessity but clearly giving an unfair advantage to the DWA, we instead use minorminer's heuristic algorithms~\cite{Cai2014} to find embeddings on a \emph{per instance} basis.

The classical postprocessing chosen for all broken chains was the majority vote algorithm. Chain strength tuning at each size for $n<80$ invariably showed that using the $extended\_j\_range$ of the device to couple logical qubits together by a $-2$ coupling strength results in the lowest TTS. This suggests that a larger ferromagnetic chain strength would be optimal.  While we have not investigated this approach in our study, one way to achieve this within the current capabilities of the hardware is to scale down the overall strength of the Ising Hamiltonian while still using a -2 coupling strength for the chain couplings. In this way, we would be able to achieve a stronger relative chain coupling. However, by reducing the overall energy scale of the problem Hamiltonian, the evolution is more susceptible to both thermal and implementation errors, so the overall rescaling of the Ising Hamiltonian would need to be optimized.

\subsection*{MemComputing}
Simulations were performed by Dr. Fabio Traversa of MemComputing Inc. on a Virtual MemComputing Machine (VMM) --- a software simulator which uses classical equations of motion to simulate the theorized Digital MemComputing Machine~\cite{Tra2015,Tra2018,Div2018}. 
DMMs are circuits of `self organizing logic gates' or (SOLGs) that can accept inputs from any of the gate terminals (either input or output) and self-organize their internal
state so that it satisfies their logic relations. These circuits are used to solve combinatorial optimization problems by
expressing the problem in boolean format and then mapping the latter onto circuits of SOLGs. 

\subsection*{SATonGPU}
In the main text we have already described the SATonGPU algorithm in some detail, and complete details can be found in Ref.~\cite{bernaschi2021leading}. 

\emph{Parallization factor.---}%
To mitigate the algorithm's dependence on initial condition Bernaschi \textit{et al.} simulated in parallel a large number of clones (or replicas), where each clone started from a different and randomly chosen initial condition. The algorithm is considered successful when any clone reaches the solution.

A total of 327680 replicas were simulated in parallel on an Nvidia V100 GPU. All runtimes reported correspond to using this many replicas. 
However, we note that the parallelization factor used in Ref.~\cite{bernaschi2021leading} appears to be weakly $n$-dependent. Nevertheless, since this amounts to at most a $\log(n)$ correction to the scaling, we did not account for it explicitly, as mentioned in the main text.

\emph{TTS extraction.---}
Ref.~\cite{bernaschi2021leading} demonstrated that the TTS, which they defined as the shortest time taken by any the replicas to find the solution of a given instance, is exponentially distributed as $P[ \text{TTS} > \tau] = \exp(-t/\tau)$.  Identifying $\tau$ is then sufficient to estimate the optTTS for SATonGPU at each size.

\begin{acknowledgements}
We are grateful to Bernaschi \textit{et al.} for sharing their data with us and to Dr. Victor Martin Mayor for clarifying discussions about the SATonGPU solver, and to Dr. Fabio Traversa for performing MemComputing simulations for us. We also thank Dr. Tamura Hirotaka and the Fujitsu team for many interesting discussions, and Mr. Yaz Izumi and the Toshiba team for feedback.
We would like to acknowledge Dr. Salvatore Mandra and Dr. Eleanor Rieffel for useful discussions and for sharing with us their benchmarking results of a separate parallel tempering implementation, which served as a useful comparison to our own. We would also like to acknowledge Dr. Aaron Lott for useful discussions concerning D-Wave devices.
The research is based upon work (partially) supported by the Office of
the Director of National Intelligence (ODNI), Intelligence Advanced
Research Projects Activity (IARPA) and the Defense Advanced Research Projects Agency (DARPA), via the U.S. Army Research Office
contract W911NF-17-C-0050. The views and conclusions contained herein are
those of the authors and should not be interpreted as necessarily
representing the official policies or endorsements, either expressed or
implied, of the ODNI, IARPA, DARPA, or the U.S. Government. The U.S. Government
is authorized to reproduce and distribute reprints for Governmental
purposes notwithstanding any copyright annotation thereon. This work is partially supported by a DOE/HEP QuantISED
program grant, QCCFP/Quantum Machine Learning and Quantum
Computation Frameworks (QCCFP-QMLQCF) for HEP, Grant No.
DE-SC0019219. The authors also acknowledge support by Fujitsu Limited, which in no way influenced any of the results reported here. 
The authors acknowledge the Center for Advanced Research Computing (CARC) at the University of Southern California for providing computing resources that have contributed to the research results reported within this publication. URL: \url{https://carc.usc.edu}.
\end{acknowledgements}

\newpage

\appendix

\section{First and third quartile optTTS results}
\label{app:A}

\begin{figure*}
    \centering
    \begin{subfigure}
    \centering
    \includegraphics[width=8.5cm]{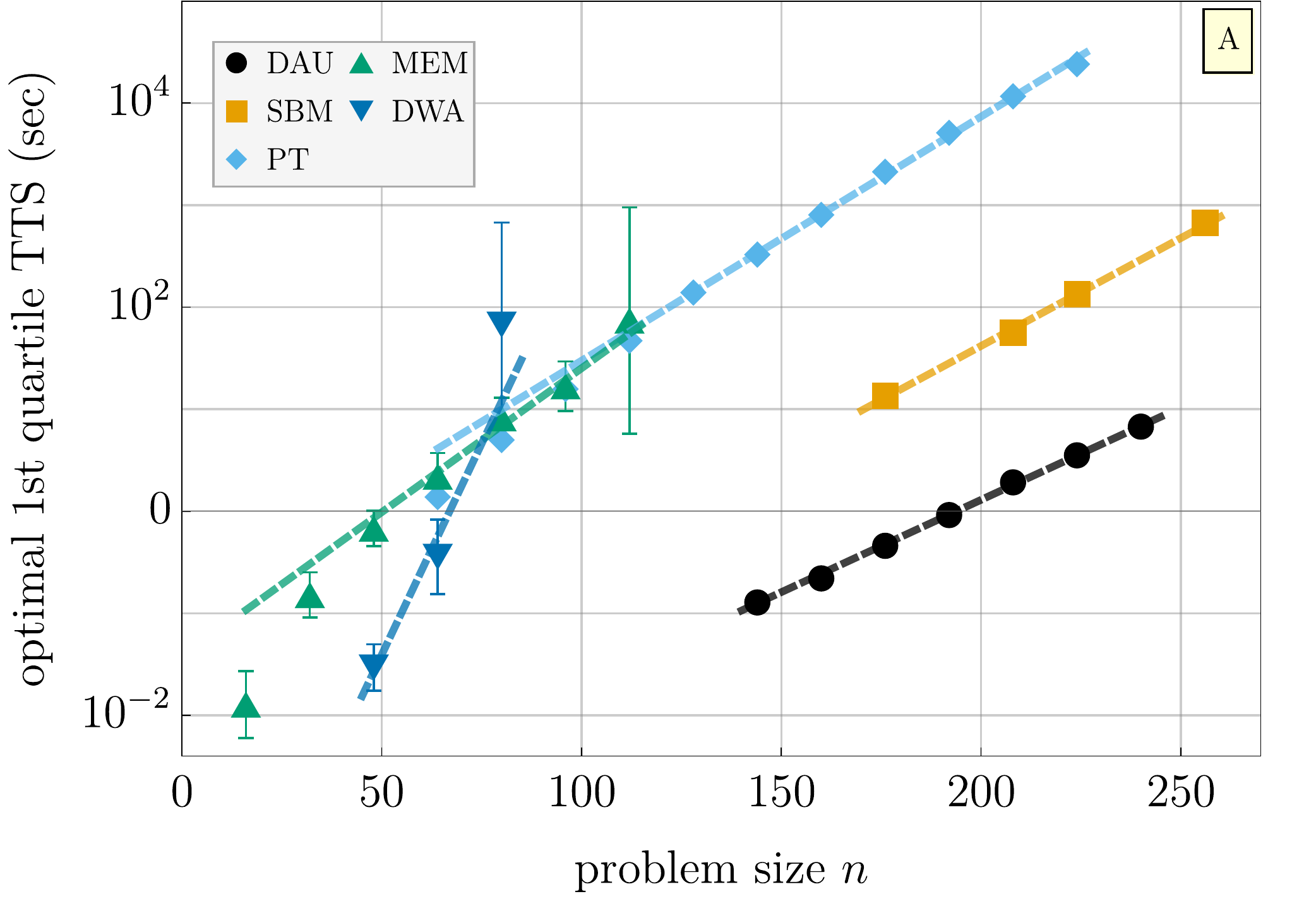}
    \end{subfigure}
    \hfill
    \begin{subfigure}
    \centering
    \includegraphics[width=8.5cm]{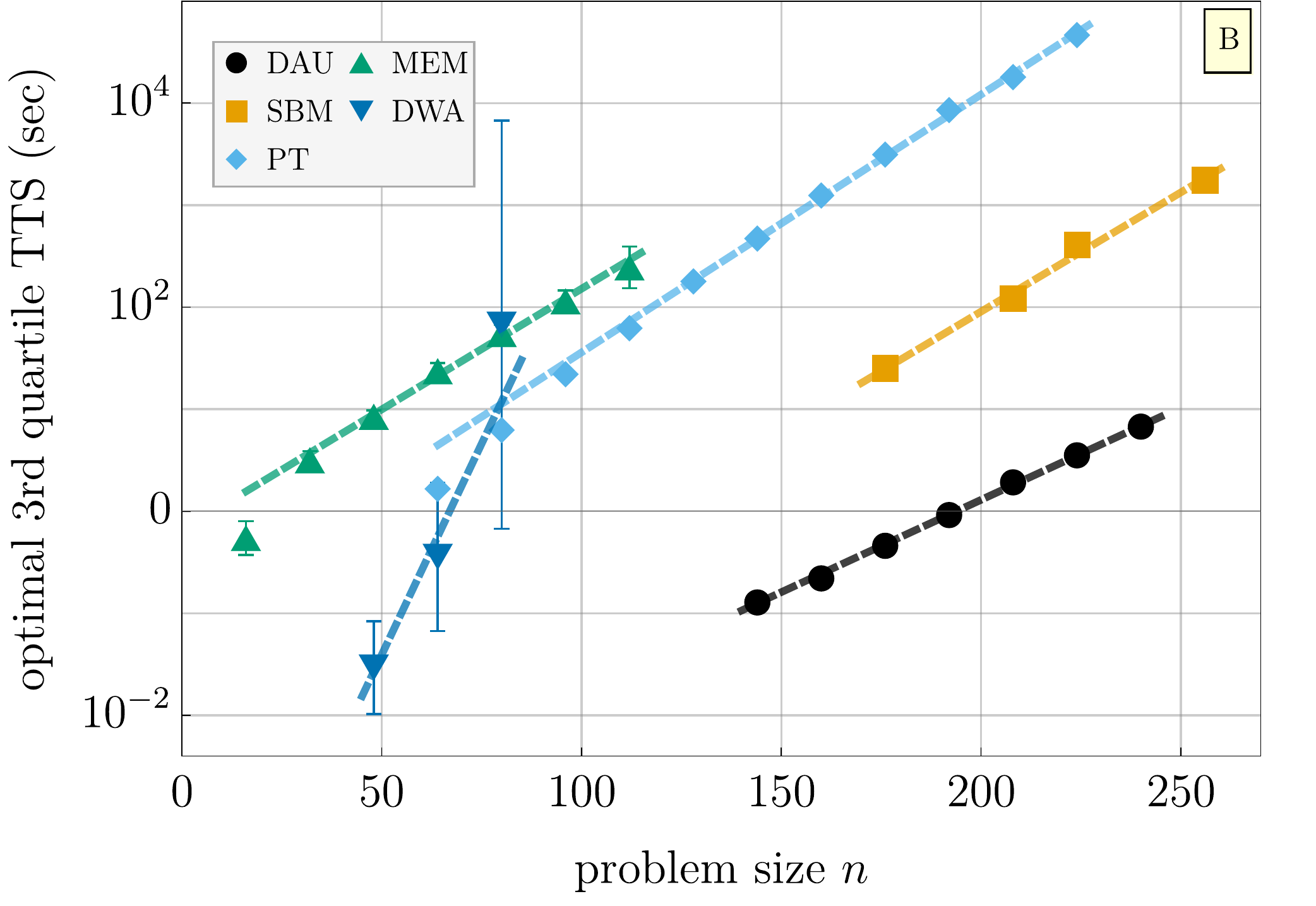}
    \end{subfigure}
    \caption{$\langle \text{TTS}\rangle_{.25}$ and $\langle \text{TTS}\rangle_{.75}$ results for 3R3X instances. Shown are the optTTS of the 25'th percentile (A) and 75'th percentile (B) hardness with 99\% probability. 
    }
    \label{fig:q25_q75}
\end{figure*}

To supplement the median results shown in Fig.~\ref{fig:3R3Xmain} in the main text, we display the first and third quartile results in Fig.~\ref{fig:q25_q75}. The corresponding fit parameters are given in Table~\ref{tab:aB_values} in the main text. These results do not alter our conclusions about the relative performance of the different solvers benchmarked in this study. 

\section{Example of the reduction from cubic form to QUBO form}
\label{app:B}

The general reduction from three-body to two-body terms is depicted graphically in Fig.~\ref{fig:3r3x_graph} of the main text.

The cost function of an $n$-bit 3R3X system is a sum of $n$ three-body terms of the form $-(-1)^{b_i} s_{i_1} s_{i_2} s_{i_3}$ defined on $n$ Ising spins where $b_i \in \{0,1\}$ and $s_{i_1},s_{i_2}$ and $s_{i_3}$ are the spins belonging to the clause.

To reduce the locality of a clause to two- and one-body terms, we use a gadget that shares its four minimizing configurations (see Ref.~\cite{Hen2019} for additional details). For the negatively signed clause ($b_i=0$), these are the four configurations whose product is positive, namely, $(1,1,1),(1,-1,-1),(-1,1,-1)$ and $(-1,-1,1)$ and an appropriate gadget is
\bea
G_{\text{3X}}&=&h(s_{i_1}+s_{i_2}+s_{i_3}) + \tilde{h} s_a \\\nonumber
&+&J(s_{i_1} s_{i_2}+s_{i_2} s_{i_3}+s_{i_3} s_{i_1})+\tilde{J} s_a(s_{i_1} +s_{i_2} +s_{i_3}) \,,
\eea
where $(h,\tilde{h},J,\tilde{J})$ can be either $(-1, -2, 1, 2)$ or $(-1, 2, 1, -2)$, yielding in both cases a minimal cost of $-4$ (we note that other gadgets exist that can equivalently be used). The above gadget is a fully connected 4-spin cost function with $s_a$ serving as an auxiliary spin. Positively signed clauses are treated similarly.

The above gadget allows us to cast $n$-bit 3R3X linear systems as two-body Ising models with $2n$ spins, as each clause adds one auxiliary spin to the problem and an $n$-bit instance has $n$ clauses.

\section{TTS parameter fitting procedure}
\label{app:C}

Our fitting procedure seeks to exclude finite size effects at small $n$ and imperfect parameter optimization at large $n$. To achieve this, beginning with two representative data points for a straight line fit on a plot of $\log_{10}\langle\text{TTS}\rangle_{q}$ \textit{vs} $n$ at intermediate $n$ values, we successively include each larger/smaller-$n$ data point until adding another data point to the fit starts to significantly alter $\alpha$. 

In order to fit the log TTS data to a linear function of the form $\alpha n + \beta$, we use Mathematica's \cite{Mathematica} built-in function `NonlinearModelFit' with the variance estimator function. We take as input to the function the logarithm of the TTS data and their $1~\sigma$ error bars; the error bars are then treated by the function as the uncertainty in the input data points. The output of the function is the best fit estimates to the coefficients $\alpha$ and $\beta$ with their $0.95$ confidence intervals, which we report in Table~\ref{tab:aB_values}.

\section{Monte Carlo time step measurement on the DAU}
\label{app:D}

Every step of the core algorithmic DAU process is executed in dedicated application specific integrated circuits (ASIC) hardware, excepting the automatic parameter adjustment for the first $10^5$ MC steps. Each hardware MC step effectively takes 52~ns. Here we explain the estimate for our parameter settings, where replica exchanges are averaged into the effective step time.

\begin{figure}[htbp]
    \includegraphics[width=\columnwidth]{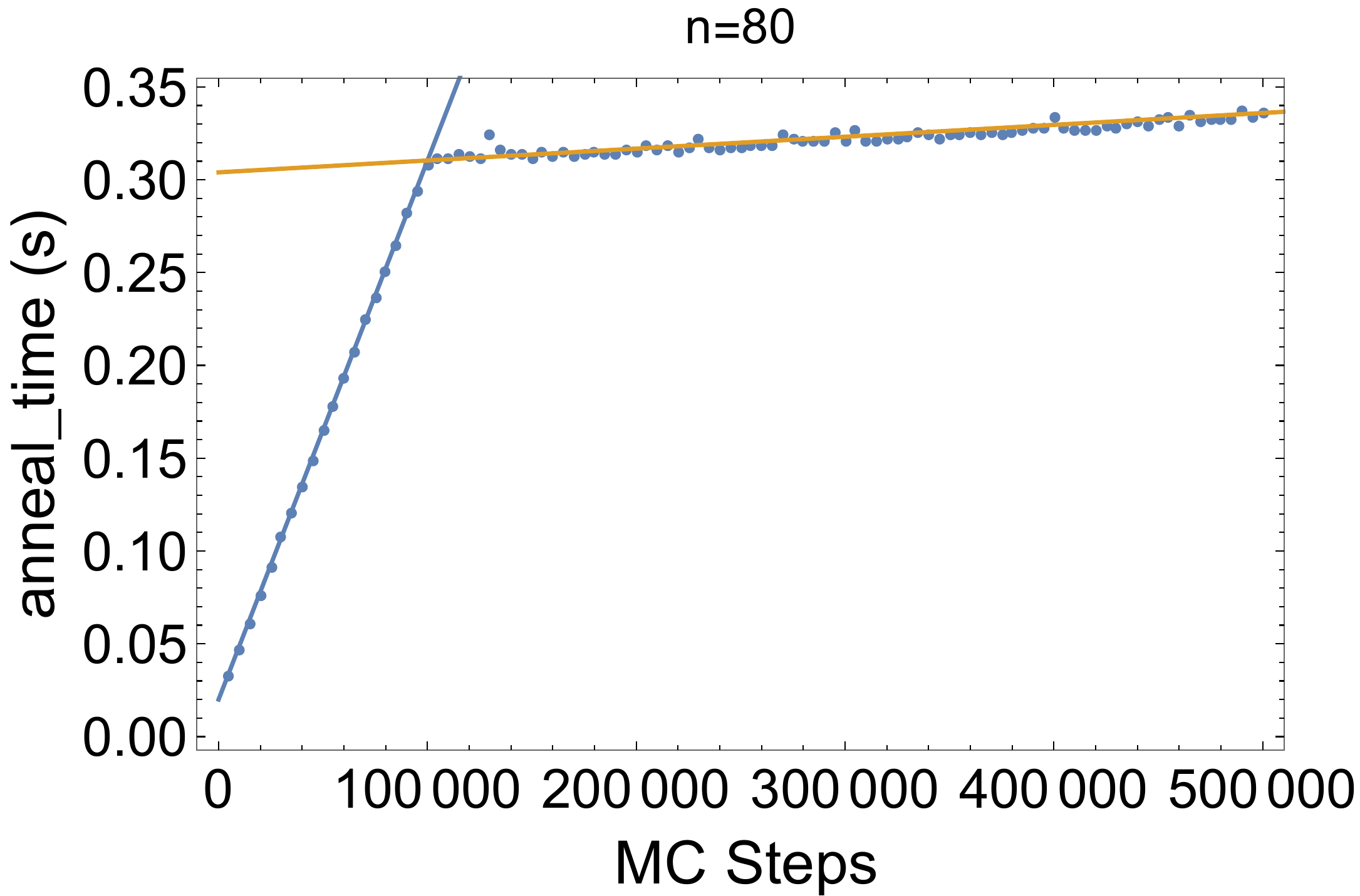}
    \caption{Extraction of the runtime per MC update $t_{MC}$ for the Fujitsu DAU from the slope of anneal time \textit{vs} number of Monte Carlo steps. Shown is a representative example with $n=80$ bits. We observe two distinct $t_{MC}$, which are the slopes seen in the plot. This holds regardless of $n$ (not shown).}
    \label{fig:twoSlopes}
\end{figure}

The DAU lacks a termination condition, i.e., it will run for the designated number of MC updates, regardless of whether it has already found the planted solution. Thus, we cannot directly extract the time-to-solution based on the returned total ``anneal\_time." 
We instead use the returned MC updates to solution $updates_{MC}$ to determine the $t_0$ defined in Methods under DAU Data Collection, which subsequently gives $p_S(t_f)$.
The needed connecting factor is the runtime per MC update $t_{MC}$, such that $updates_{MC}*t_{MC} = t_0$. 
Many runs were performed for a total number of MC updates at intervals of $repex\_interval=2500$, then post-selected only for runs which did not reach the solution state and averaged. A line fit to this data has a slope of effective $t_{MC}$. However, there are two distinct values of $t_{MC}$ as illustrated for a representative case in Fig.~\ref{fig:twoSlopes}.
For the first 100,000 MC steps, $t_{MC}\approx 3 \times 10^{-6}$s, then for all subsequent MC updates $t_{MC} \approx 52 \times 10^{-8}$s. This finding is consistent regardless of problem size $n$, and derives from the use of automatic parameter adjustment on a CPU for the first 100,000 MC steps. To find the scaling associated with purely the algorithm, we use $t_{MC}=52$ns for calculating TTS in the main text.

\section{D-Wave Detailed Data Collection}
\label{app:E}

We detail the procedure used in our study for data collection from the DWA. We used the Ising form of the 3R3X instances rather than the QUBO form.

\begin{enumerate}
    \item Use Minorminer.find$\_$embedding with the following parameter setting: (max\_no\_improvement=100, chainlength$\_$patience=100, tries=100, timeout=1000(s)), to find and store \emph{per instance} embedding.
    \item Generate and save at least 50 gauges of each Ising problem.
    \item Choose a subset of 10 instances and 6 gauges at each $N$ to run for log-uniform distribution of 11 annealing times from $1 \mu s$ to $2000 \mu$s and the following set of chain strengths: (2,1.5,1).
    \item Randomize the instance/gauge list.
    \item For each instance/gauge:
    \begin{enumerate}
        \item Load the corresponding embedding, use VirtualGraphComposite
        \item Manually rescale the Ising problem and its ground state (GS) energy to exactly fit the bias/coupler ranges of the solver.
        \item Randomize chain strength list and annealing time list.
        \item For each chain strength/annealing time combination:
        \begin{enumerate}
            \item Sample $250-500$ times, depending on $n$.
            \item Store the full set of raw data.
            \item Use majority vote decoding to ``fix" broken chains, store percentage of broken chains.
            \item Store a lightly processed data set with number of successes derived from comparison with scaled GS energy.
        \end{enumerate}
    \end{enumerate} 
    \item For each $n$  and chain$\_$strength set of data perform an optTTS analysis over annealing times that bootstraps over instances/gauges in the following way at each annealing time:
    \begin{enumerate}
        \item For each of $\ge 1000$ bootstraps do:
        \begin{enumerate}
            \item resample $I$ instances
            \item for each instance:
            \begin{enumerate}
            \item set $numSuccess = numFailures=0$.
            \item resample $G$ gauges
            \item For each gauge add the number of successful runs (that found the GS) to $numSuccess$, and add the number of failed runs to $numFailures$ (\emph{do not} sample $p_S$ here).
            \end{enumerate}
            \item Sample $p_S$ from the following distribution: $\beta(numSuccess+0.5,numFailures+0.5)$, store in $p_S$ vector.
        \end{enumerate}
        \item Calculate $p_{Sq}$ here by finding the $q$th percentile $p_S$ of the current instance's $p_S$ vector of data, store $p_{Sq}$.
    \end{enumerate}
    \item After all $\ge 1000$ bootstraps are complete, store the mean/standard deviation of across $p_{Sq}$'s in the $p_{Sq}$ vector over instances to produce the mean TTS at that annealing time.
    \item For each $n,chain\_strength$ combination, estimate the minimum TTS across annealing times via a fitting function, as illustrated in Fig.~\ref{fig:TTS fitting}.
        \begin{figure}
        \centering
        \includegraphics[width=\columnwidth]{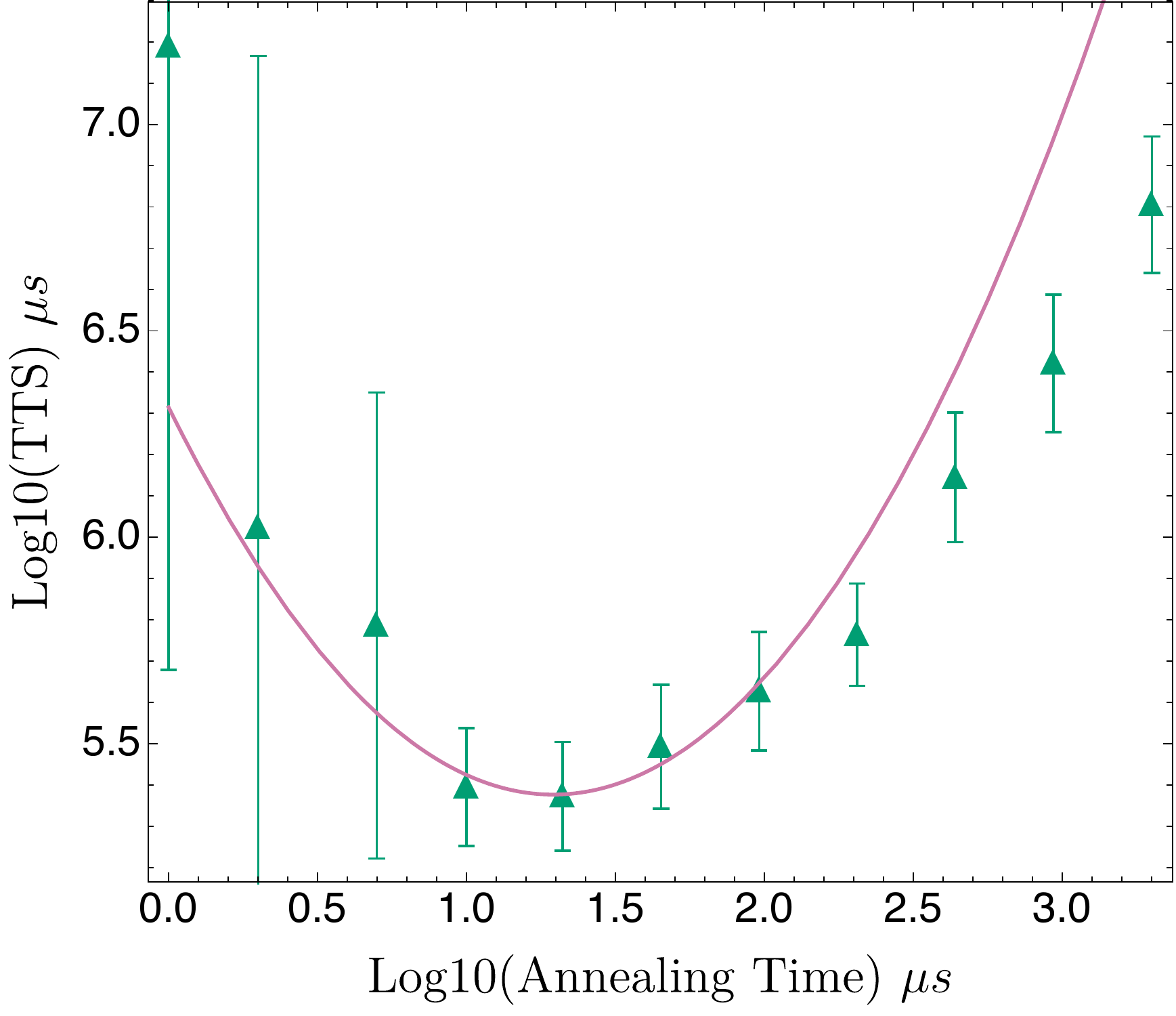}
        \caption{An example of fitting the TTS \textit{vs} annealing time for $n=64$ to find  optTTS and the optimal annealing time for our DWA analysis.}
        \label{fig:TTS fitting}
    \end{figure}
    \item Finally, select the best $chain\_strength$ value at each $n$ and plot in the final comparison with other solvers.
\end{enumerate}

\section{Toshiba SBM Origins}
\label{app:F}

\emph{Background.---}% 
Networks of Kerr Parametric Oscillators (KPOs) were proposed recently as a method of quantum adiabatic optimization (QAO) for Ising/QUBO problems~\cite{Got2016,Pur2017}. Following Goto~\cite{Got2016}, consider the Hamiltonian
\begin{align}
H_q(t)&=\hbar \sum_{i=1}^{n}\left[\frac{K}{2} a_{i}^{\dagger 2} a_{i}^{2}-\frac{p(t)}{2}\left(a_{i}^{\dagger 2}+a_{i}^{2}\right)+\Delta_{i} a_{i}^{\dagger} a_{i}\right] \nonumber \\ 
&\qquad\qquad -\hbar \xi_{0} \sum_{i<j}^{n} J_{ij} a_{i}^{\dagger} a_{j} \ , \label{eq:sbm_Hq}
\end{align}
where $\hbar$ is the reduced Planck constant, $a_i^\dagger$/$a_i$ is the creation/annihilation operator for the $i$'th oscillator, $K$ is a positive Kerr coefficient, $p(t)$ is the time-dependent parametric two-photon pumping amplitude, $\Delta_i$ is the positive detuning frequency between the resonance frequency of the $i$'th oscillator and half the pumping frequency, and $\xi_0$ is a positive constant with dimensions of frequency. Each KPO is initially set to the vacuum state, $p(0)$ is set to 0, and $\xi_0$ is set small enough that the vacuum state is the ground state of the initial Hamiltonian. If by the final time $t_f$ the pumping amplitude is sufficiently large ($p(t_f) \gg \max_{i}\Delta_i$), then each KPO reaches a coherent state with a positive or negative amplitude via a quantum adiabatic bifurcation. If $p(t)$ is increased slowly enough, the quantum adiabatic theorem guarantees the sign of the final amplitude for the $i$'th KPO provides the $i$'th Ising spin of the ground state of the Ising problem~\cite{Got2016}.

\emph{Algorithm.---}% 
While a physical realization of a network of quantum KPOs for QAO has yet to be realized, a new classical algorithm for Ising/QUBO problems derived from a classical approximation of  Eq.~\eqref{eq:sbm_Hq} has been proposed and demonstrated~\cite{Got2019b}. By approximating $a_j = x_j + iy_j$, a classical Hamiltonian and derived equations of motion are obtained. Based on empirical observations~\cite{Got2016, Got2019} of the $y_j$ amplitude oscillating around $0$ and the need for fast numerical simulation, the authors remove some terms linear in $y_j$ from the equations of motion. Assuming all detunings have the same value $\Delta$, they obtain the final equations used in the simulated bifurcation (SB) algorithm:
\bes
\begin{align}
\dot{x}_{i}&=\Delta y_{i} \label{eq:SB1}\\ 
\dot{y}_{i}&=-\left[K x_{i}^{2}-p(t)+\Delta\right] x_{i}+\xi_{0} \sum_{j=1}^{N} J_{ij} x_{j}\ . \label{eq:SB2}
\end{align}
\ees
The SB algorithm is as follows. Use the constant values determined by the KPO-AQO method with the normalization $K=\Delta=1$ and set all initial $x_i$ and $y_i$ around zero. Gradually increasing $p(t)$ from zero, solve Eqs.~\eqref{eq:SB1} and~\eqref{eq:SB2} numerically by a modified explicit symplectic Euler method (See Methods in~\cite{Got2019b}). The sign of the final $x_i$ provides that of the $i$'th spin of an approximate solution of the Ising problem.

%\bibliography{References}

%%%%%%%%%%%%%%%%%%%%%%%%%%%%%
\end{document}